\documentclass[%
 reprint,
 amsmath,amssymb,
 aps,
]{revtex4-2}

\usepackage{graphicx}% Include figure files
\usepackage{dcolumn}% Align table columns on decimal point
\usepackage{bm}% bold math
\usepackage{newunicodechar}
\newunicodechar{∥}{\ensuremath{\parallel}}
\begin{document}

\preprint{APS/123-QED}

\title{A Flux-Tunable Discrete Angular Filter}% Force line breaks with \\

\author{Tristan Lawrie}

 \email{t.lawrie@exeter.ac.uk}
 
\affiliation{Centre for Material Research and Innovation, Department of Physics and Astronomy, University of Exeter}

\affiliation{School of Mathematical Sciences, University of Nottingham}

\author{Oliver M. Brown}
\affiliation{School of Mathematics, University of Bristol}



\date{\today}% It is always \today, today,

\begin{abstract}
Recent work by Lawrie et al. [PRR 7, 023209 (2025)] introduced a non-diffracting resonant angular filter on a network of thin channels (modelled via quantum graph theory) that exhibits unit transmission of acoustic waves at a discrete, symmetry-paired set of incidence angles determined solely by the graph topology, while transmission at all other angles is strictly forbidden. In the present work, we study the same filtering geometry for waves governed by the magnetic Schrödinger equation rather than the classical wave equation. Using a phase shift induced by non-reciprocal wave propagation due to the presence of the magnetic potential and tuning $\delta$-type vertex boundary conditions, we make the previously topology-fixed discrete pass directions continuously tunable: both the transmission angle and the transmission coefficient become control parameters. The resulting flux-tunable angular filtering device replaces topology-constrained passbands with a programmable steering device, broadening the scope for wave-filter and beam-shaping applications.

% \ell

% X

% Y

% Z

% $m$

% $m + \mu$

% $m - \mu$

% $a_{m,r}^{\text{in}}$ $a_{m,r}^{\text{out}}$

% $a_{m,l}^{\text{in}}$ $a_{m,l}^{\text{out}}$

% $a_{m,d}^{\text{in}}$ $a_{m,d}^{\text{out}}$

% $a_{m,u}^{\text{in}}$ $a_{m,u}^{\text{out}}$

% $r$

% $1$ $r_{\mu}$ $t_{\mu}$

% $\theta$

% $r_{\mu}$

% (a)

% (b)

% (c)

% $R$

% $(a)$

% $(b)$

% $(c)$

% $(d)$

% $\theta$

% $|t_{\mu}(\theta)|^2$

% $\Delta \Phi$

% $\Delta \lambda$

% $n$

% $-95$

% $0$

% $95$

% $m$

% $-120$

% $0$

% $120$

% $\mathbb{R}\{\psi_{n,m,e}\}$

% $-0.50$

% $-0.25$

% $0.00$

% $0.25$

% $0.50$

\end{abstract}

\maketitle

%\tableofcontents

\section{Introduction}\label{sec:introduction}

The filter in this paper extends the non-diffracting acoustic resonant angular filter from Lawrie et al. \cite{lawrie2025nondiffracting}. The filter gives unit transmission at discrete symmetrically paired angles, while transmission for all other angles is strictly forbidden. The filter's geometry mimics that of a grating \cite{wilcox2012scattering}, with periodic openings connected not only to the scattering environments but also internally via thin channels of variable length. This allows connections from opening $m \in \mathbb{Z}$ to $m \pm \mu$ for $\mu \in \mathbb{N}$, forming beyond-nearest-neighbour connections first shown by Brillouin \cite{Brillouin1960}. Such metamaterials \cite{KUMAR20223016} enable non-local interference and Roton-like dispersion \cite{Wang2024, chen2021roton, wang2022nonlocal, iglesias2021experimental, chaplain2023reconfigurable, moore2023acoustic, chen2023observation, kazemi2023non, fleury2021non, kazemi2023drawing}, yielding non-trivial wave properties. In the current work, the value of $\mu$ informs the number of unit transmission angles. The discrete filtering properties arise from a resonance where internal channels support a harmonic mode decoupling from the environment, known as bound states in the continuum \cite{lawrie2023closed}. Under resonance, junctions impose effective Dirichlet boundary conditions that block mode coupling — see \cite{lawriethesis} for a detailed analysis. At specific angles, the incident field's tangential wave number aligns with the filter's bound state, switching conditions from Dirichlet to periodic, allowing for unit transmission. Similar effects have been analysed in the frequency domain in \cite{DAB19, DAB20, akhshani2023quantum}. In the present work, we enhance the filter's tunability by adding a magnetic potential to internal channels and varying transmission amplitude via $\delta$-type boundary conditions at junctions. This results in a flux-tunable angular filtering device.

Transmission properties are analysed using the scattering language of quantum graph theory, first applied to metamaterials in \cite{lawrie2022quantum} and validated against COMSOL simulations \cite{lawrie2025nondiffracting} and acoustic experiments \cite{lawrie2024application}. This theory examines spectra of graph structures with one-dimensional self-adjoint differential operators and variable junction conditions \cite{berkolaiko2013introduction, kottos1999periodic}, as the thin-channel limit of networks \cite{kuchment2001convergence, rubinstein2001variational, exner2005convergence}. Originally formulated for the study of the Schrödinger equation, it applies to quantum chaos \cite{gnutzmann2006quantum, kottos1999periodic, gnutzmann2006quantum}, quantum random walks \cite{kempe2003quantum, tanner2006quantum}, and lattice wave communication \cite{hein2009wave}. It extends to other operators, including the wave equation \cite{lawrie2024application}, linearized Korteweg-de Vries equation \cite{smith2025linearized}, and beam/plate equations \cite{edge2025discrete, brewer2018elastodynamics}. But most relevant to this work is the magnetic Schrödinger equation \cite{Gnutzmann__2006, Popov2020}, where magnetic potential shifts effective wave numbers non-reciprocally for forward/backward propagation, breaking time-reversal symmetry. This formalism applies to analogous non-reciprocal wave transport contexts, such as non-reciprocal optical waveguides~\cite{JIAO2025,Shoji01022014,PhysRevLett.104.253902,6602935,dnp4-z1xk,Shen2016,Kodz:24,Tzuang2014,Shoji2016,Yokoi:99} and acoustic crystals with magnetochiral phonon propagation~\cite{PhysRevLett.122.145901}. Additionally such effects have been shown in time-varying metamaterials where the breaking of time reversal symmetry can be achieved via the Doppler effect \cite{caloz2019spacetime}.

The paper is structured as follows: In Section \ref{sec: Modelling the Filter via Quantum Graph Theory} We determine the scattering properties of the filter in terms of a quantum graph model and derive an analytic equation for the transmission coefficients of the graph influenced by a magnetic potential. In Section \ref{sec: results} the resulting scattering properties of the filter are analysed and demonstrated numerically for a variety of filter configurations. Finally in Section \ref{sec: Conclusion} this work is concluded. 

\section{Modelling the Filter via Quantum Graph Theory}\label{sec: Modelling the Filter via Quantum Graph Theory}

\begin{figure*}[t!]
    \centering
    \includegraphics[width=1\textwidth]{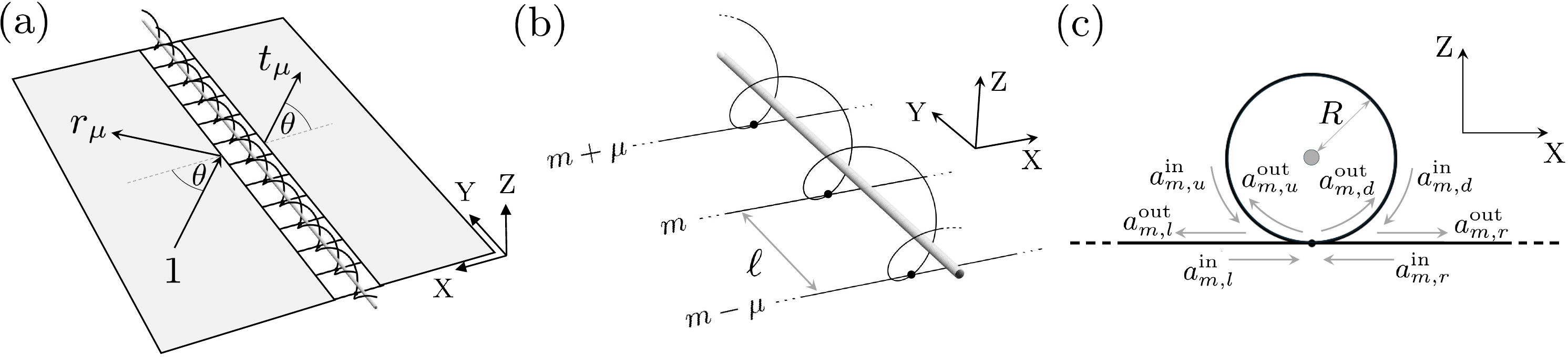}
    \caption{(a) shows a schematic of the filter placed between two semi-infinite half spaces (shown in grey), where waves are free to travel. Illustrated via an arrow is a wave of amplitude $1$ with incidence angle $\theta$ scattering at the filter into a reflected wave with amplitudes $r_{\mu}$ and a transmitted wave with amplitude $t_{\mu}$. (b) shows a schematic of the filter itself which is formed of an array of vertices with period $\ell$ and discrete index $m$ coupled to the scattering environment by leads to the left ($l$) and right ($r$) of the vertex and to one another in the down ($d$) and up ($u$) directions via a bond of length $\ell_{\mu}$, here $\mu = 1$. The connected bonds form a helix, through which a solenoid is placed which will induce a magnetic potential through the bonds. (c) shows a side view of the filter where the solenoid is a constant distance $r$ away from the graph bonds; emphasised are the local edge wave amplitudes heading in and out of vertex $m$.}
    \label{fig: Filter Schematic}
\end{figure*}

The aim of this section is to determine the reflection $r_{\mu}$ and transmission $t_{\mu}$ coefficients of the preposed filter, illustrated in Fig.~\ref{fig: Filter Schematic} (a). The filter, modelled as a metric graph $\Gamma(\mathcal{V},\mathcal{E},L)$ embedded in $\mathbb{R}^3$, is constructed from an infinite set of vertices $\mathcal{V}$ placed along the $Y$-axis with period $\ell$ and discrete index $m\in\mathbb{Z}$, coupled by an infinite set of bidirectional edges $\mathcal{E}$ each with metric length $L = \{l_e \in \mathbb{R}^+ \mid e \in \mathcal{E} \}$. The set of edges with finite length will be called bonds $\mathcal{B}$, while the set of edges with semi-infinite length will be called leads $\mathcal{L}$, with union $\mathcal{E}=\mathcal{L} \cup \mathcal{B}$.
Each vertex $m$ is coupled up ($u$) and down ($d$) to vertices $m\pm \mu$ for $\mu \in \mathbb{N}$ by bonds of length $\ell_u = \ell_d :=\ell_\mu$. The graph is made open by coupling leads along the $x$-axis to the left ($l$) and right ($r$) of the vertex which will later be coupled to the scattering environment as in \ref{fig: Filter Schematic} (a). The set of edges coupled to a given vertex $m$ form a sub set of $\mathcal{E}$ which we define as the star of the vertex $\mathcal{S}_m= \{l_m,r_m,d_m,u_m\}$. For each edge $e\in\mathcal{S}_m$, we introduce an edge coordinate $z_{m,e}$ with origin chosen to be at vertex $m$ which spans the domain $z_{m,l} \in [0,\infty)$, $z_{m,r} \in [0,\infty)$ and $z_{m,d} \in [0,\ell_\mu]$, $z_{m,u} \in [0,\ell_\mu]$. Note the use of $z$ as an edge coordinate rather than the traditional Euclidean coordinates $X,Y$ and $Z$.
We emphasize that in this description, each edge in the filter admits \textit{two} parameterizations, $z_{m,u/d}$ with the vertex labelled $m$ at the origin and $z_{m\pm\mu,d/u}$ with vertex $m$ located at $\ell_{\mu}$. These parameterizations satisfy $z_{m \pm \mu, d/u} = \ell_{\mu} - z_{m, u/d}$

The metric graph is turned into a quantum graph by imposing a wave equation on each edge, as well as enforcing a choice of boundary conditions on each vertex. Here we consider the one-dimensional magnetic Schrödinger equation,
\begin{equation}\label{eq:schrodinger}
    \left( - i \frac{\partial}{\partial z_{m,e}} + A_{m,e} \right)^2 \psi_{m,n} (z_{m,e}) = k^2 \psi_{m,e} (z_{m,e}),
\end{equation}
where the real constant $A_{m,e}$ represents a magnetic potential along edge $e$ connected to vertex $m$. This term can be introduced physically into the system in a myriad of ways, but for this work we consider the bonds to be wrapped around a solenoid forming a helix like-structure of constant radius $r$ as shown clearly in Fig. \ref{fig: Filter Schematic} (b) and (c). We assume that the solenoid is suitably far enough away from the leads $e = l_m/r_m$ and equal distance from all vertices such that $A_{m,l} = A_{m,r} = 0$ and $A_{m,d} = - A_{m,u}$. The general solution of equation (\ref{eq:schrodinger}) is a superposition of counter-propagating plane waves:

\begin{equation}\label{eq:schrodingersolution}
    \psi_{m,e} = \text{e}^{i(\kappa_ym\ell - A_{m,e}z_{m,e})} \left( a^{\text{out}}_e \text{e}^{i k z_{m,e}} + a^{\text{in}}_e \text{e}^{- i k z_{m,e}} \right).
\end{equation}

Here, $a^{\text{out/in}}_e$ are the complex wave amplitudes heading out of or into a given vertex on edge $e$ as illustrated in Fig. \ref{fig: Filter Schematic} (c). The phase term $\text{e}^{i\kappa_y m \ell}$ is the Bloch phase \cite{kittel2005introduction}, which arises as a consequence of the lattice periodicity. The Bloch wave number $\kappa_y$ defines the angle $\theta$ of a wave incident on the filter. The effective wave numbers $k-A_{m,e}$ and $k+A_{m,e}$ induce different momenta in the waves that propagate along each edge. This leads to non-reciprocal wave transport thus breaking time-reversal symmetry.

Next, we consider the vertex boundary conditions that couple waves on different edges. In principle, one can consider a wide range of options, including those of resonant cavities \cite{aldosri2024wave}, those that break time-reversal symmetry \cite{baradaran2024cairo}, or the most general forms that preserve self-adjointness \cite{KSchr99}. One could even forgo self-adjointness entirely, such that energy is not conserved as with leaky vertices. Here, however, we focus on $\delta$-type boundary conditions, since they allow us to tune the filter's transmission coefficient using a single parameter. Such $\delta$-type boundary conditions must satisfy the following conditions:
\begin{enumerate}
    \item the wave functions on connected edges $e$ and $e'$ are continuous at their shared vertex $v_m$, parameterized at $0$,\begin{equation}\label{eq:continuity}
        \psi_{m,e} (0) = \psi_{m,e'} (0);
    \end{equation}
    \item the outgoing momenta of the wave function on each adjacent edge $e$ to vertex $m$ is given as a constant function~\cite{Gnutzmann__2006},
    \begin{equation}\label{eq:kirchoff}
        \sum_{e \in \mathcal{S}_m} \left( \frac{\partial }{\partial z_{m,e}} - i A_{m,e} \right) \psi_{m,e} (0) = \lambda \psi_{m,e} (0).
    \end{equation}
\end{enumerate}
Here $\lambda$ is a free parameter that we choose for an ideal transmission profile. We introduce the gauge transformation
\begin{equation}
    \psi_{m,e} (z_{m,e}) \rightarrow \text{e}^{i \int_{0}^{z_{m,e}} A_{m,e} d z_{m,e}} \psi_{m,e} (z_{m,e}),
\end{equation}
integrating from the vertex with metric position $0$ on $e$ to $z_{m,e}$, the potential can be removed from (\ref{eq:schrodinger}) so that the waves propagate with equal and opposite momenta.
The new vertex conditions for the gauge-transformed wave function $\phi_{m,e}$ are
\begin{equation}\label{eq:gaugetransformedcontinuity}
    \phi_{m,e} (0) = \phi_{m,e'} (0)
\end{equation}
from (\ref{eq:continuity}), and
\begin{equation}\label{eq:gaugetransformedkirchoff}
    \sum_{e \in \mathcal{S}_m} \frac{\partial \phi_{m,e}}{\partial z_{m,e}} (0) = \lambda \phi_{m,e} (0)
\end{equation}
from (\ref{eq:kirchoff}), achieving the conventional Kirchoff-Neumann conditions at $v_m$ in the parameterization of edges that has $v_m$ at the origin; however, we note that when comparing the values of the gauge-transformed wave function parameterized between $v_m$ and $v_{m \pm \mu}$ we have
\begin{equation}\label{eq:aharonovbohm}
        \text{e}^{ i \int_{0}^{\ell_{\mu}} A_{m,e} d z_{m,e}} \phi_{m,e} (\ell_{\mu}) = \phi_{m \pm \mu, e} (0),
\end{equation}
which transfers the action of $A_e$ to a phase change in waves that propagate around the solenoid in a manner analogous to the \textit{Aharonov-Bohm} effect. Through substitution of equation (\ref{eq:schrodingersolution}) into (\ref{eq:gaugetransformedcontinuity}) and (\ref{eq:gaugetransformedkirchoff}) it is trivial to show that the transmission $\tau$ and  reflection $\rho$ coefficients between edges at a joining vertex are given as, 
\begin{align}\label{eq: Vertex Transmission and Reflection}
\tau(k;\lambda) = \frac{2ik}{vik - \lambda} &&
\rho(k;\lambda) = \tau - 1.
\end{align}
where $v = |\mathcal{S}_m| = 4$ is the valency (number of connected edges) at the vertex. See a line-by-line derivation of these coefficients in  \cite{lawriethesis}. Note the limits restore effective vertex boundary conditions,
\begin{equation}
    \lambda = 
    \begin{cases}
        0, & \text{Neumann-Kirchhoff boundary conditions}, \\
        \infty, & \text{Dirichlet boundary conditions}.
    \end{cases}
\end{equation}
With that, we have everything needed to determine the scattering properties of the filter.

\begin{figure*}[t!]
    \centering
    \includegraphics[width=1\textwidth]{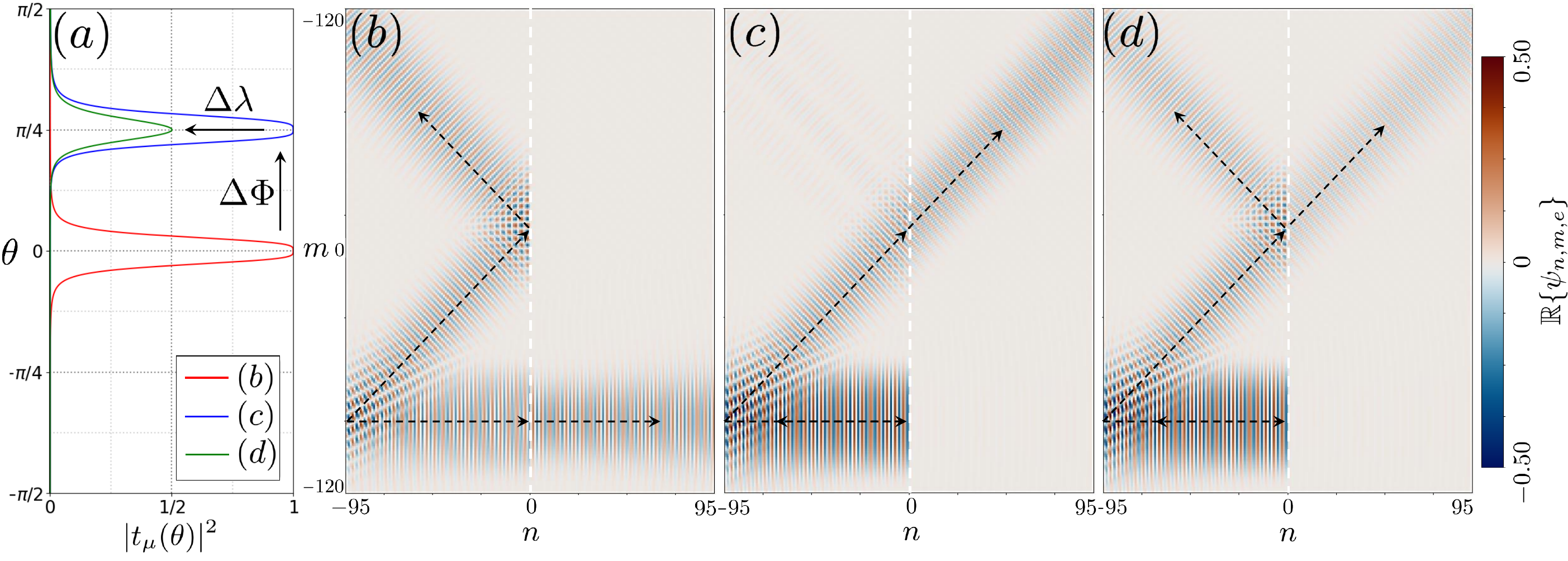}
    \caption{(a) The squared magnitude of the transmission coefficient from Eq.~(\ref{eq: Transmission Definition}) for a filter with unit period \(\ell = 1\), nearest-neighbor vertex connections (\(\mu = 1\)), and bond length \(\ell_{\mu} = 2\pi/k + \epsilon\) where \(\epsilon = 0.01\) (off-resonance). The red curve corresponds to \(\Phi = \lambda = 0\), showing unit transmission at zero angle. The blue curve illustrates the effect of changing the magnetic potential to \(\Phi = 1\), while the green curve shows the effect of changing the vertex parameter to \(\lambda = 2\). (b--d) Effects of different boundary configurations on two incident beams at angles $\theta = 0$ and \(\pi/4\) radians. (b) Unit transmission is achieved for the zero-angle beam, while transmission is excluded for the beam at incident angle \(\theta = \pi/4\). (c) Unit transmission is achieved for the beam at incident angle \(\theta = \pi/4\). (d) The transmission coefficient is chosen such that an ideal incident angle is split evenly at the boundary.}
    \label{fig: Results}
\end{figure*}

Before we go on, a bit of book keeping is required, given the large number of wave amplitudes we need to keep an eye on. As such we now sort out ingoing and outgoing wave amplitudes on each bond $\mathcal{B}$ and lead $\mathcal{L}$ into vectors $\mathbf{a}_{\mathcal{B}}^{\text{out/in}} = \left(a_{d}^{\text{out/in}}, a_{u}^{\text{out/in}} \right)^T$ and $\mathbf{a}_{\mathcal{L}}^{\text{out/in}} = \left(a_{l}^{\text{out/in}}, a_{r}^{\text{out/in}} \right)^T$ respectively, and relate them via the scattering matrices,
\begin{equation}\label{eq: Scattering 1}
\mathbf{a}_{\mathcal{L}}^{\text{out}} = S_{\mathcal{LL}}\mathbf{a}_{\mathcal{L}}^{\text{in}} + S_{\mathcal{LB}}\mathbf{a}_{\mathcal{B}}^{\text{in}}
\end{equation}
and
\begin{equation}\label{eq: Scattering 2}
\mathbf{a}_{\mathcal{B}}^{\text{out}} = S_{\mathcal{BL}}\mathbf{a}_{\mathcal{L}}^{\text{in}} + S_{\mathcal{BB}}\mathbf{a}_{\mathcal{B}}^{\text{in}}.
\end{equation}
Where the $S_{\mathcal{LL}}$ is the scattering matrix that maps wave amplitudes between leads, $S_{\mathcal{LB}}$ is the scattering matrix that maps wave amplitudes between bonds and leads, $S_{\mathcal{BL}}$ is the scattering matrix that maps wave amplitudes between leads and bonds and $S_{\mathcal{BB}}$ is the scattering matrix that maps wave amplitudes between bonds. Each matrix is constructed from the vertex scattering coefficients $\tau$ and $\rho$ as,
\begin{equation}
S_{\mathcal{LL}} = S_{\mathcal{BB}} = \tau \mathbb{E}_{2} - \mathbb{I}_{2}
\end{equation}
and
\begin{equation}
S_{\mathcal{LB}} = S_{\mathcal{LB}} = \tau \mathbb{E}_{2}.
\end{equation}
Of course $\mathbb{E}_{2}$ represents a matrix of all ones of dimension $2$ and $\mathbb{I}_{2}$ is the identity matrix of equivelent dimension. By evaluating the phase dynamics along each bond, we write a phase matrix $P$ that performs the mapping,
\begin{equation}\label{eq: Phase Matrix}
\mathbf{a}^{\text{in}}_{\mathcal{B}} = P(k)\mathbf{a}^{\text{out}}_{\mathcal{B}}
\end{equation}
By combining equations (\ref{eq: Scattering 1}), (\ref{eq: Scattering 2}) and (\ref{eq: Phase Matrix}) we determine the lead scattering matrix $S_{\mu}$ and thus the scattering properties of the filter. Here $S_{\mu}$ performs the mapping
\begin{equation}
\mathbf{a}^{\text{out}}_{\mathcal{L}} = S_{\mu}\mathbf{a}^{\text{in}}_{\mathcal{L}},
\end{equation}
where,
\begin{equation}\label{eq: Filter Scattering Matrix}
S_{\mu} = S_{\mathcal{LL}} + S_{\mathcal{LB}}\left[\mathbf{I} - PS_{\mathcal{BB}}\right]^{-1}PS_{\mathcal{BL}}.
\end{equation}
Here we are left to define $P$. For this we consider the particle that defines the wave function to have unit charge and the vector potential $\mathbf{A}$ induces on the particle by the solenoid is, in the Coulomb gauge
\begin{equation}
    \mathbf{A} = \frac{\Phi}{2 \pi R} \hat{\Theta}
\end{equation}
written in cylindrical coordinates $(X,Y,Z)\rightarrow(R,Y,\Theta)$, and with $\Phi$ equal to the magnetic flux through the solenoid. We express the bond coordinated in terms of the Euclidean axis ($X,Y,Z$) rather than the edge axis $z_{m,e}$ and parametrize $\mathbf{r}_u:[0,\ell_{\mu}]\rightarrow\mathbb{R}^3$ in cylindrical coordinates $\left( R, Y , \Theta \right)$ as
\begin{equation*}
    \mathbf{r}_u (s) = \left( R, \frac{2 \pi s}{\ell_{\mu}}, s\sqrt{1 - \left( \frac{2 \pi R}{\ell_{\mu}} \right)^2 }\right)^T,
\end{equation*}
on edge $u$ relative to $v_m$ and $\mathbf{r}_{d}(s) = \mathbf{r}_{u}(\ell_{\mu} - s)$.
The integral of $\mathbf{A}$ along the bond is always equal to simply the flux enclosed,
\begin{equation}
    \int_{u} \mathbf{A} \cdot \mathbf{dr}_u  = \int_{0}^{\ell_{\mu}} A_{m,u} d z_{m,u} = \int^{\ell_\mu}_{0} \frac{\Phi}{2 \pi R} \cdot \frac{2 \pi}{\ell_{\mu}} \, R ds = \Phi
\end{equation}
and consequently the gauge-transformed wave function must satisfy the conditions
\begin{equation*}
    e^{i \Phi} \phi_{m,u} (\ell_{\mu}) = \phi_{m+\mu,d} (0),
\end{equation*}
and
\begin{equation*}
    e^{- i \Phi} \phi_{m,d} (\ell_{\mu}) = \phi_{m-\mu,u} (0),
\end{equation*}
that equate wave functions parameterized from different vertices, as in equation (\ref{eq:aharonovbohm}).
These conditions can be written as a matrix equation in the amplitudes.
\begin{equation}\label{eq:transfer}
    \mathbf{a}_{\mathcal{L}}^{\text{in}}
    = e^{i k \ell_{\mu}} \begin{pmatrix}
        0 & e^{i \Phi} e^{- i \kappa_y \mu \ell} \\
        e^{- i \Phi} e^{i \kappa_y \mu \ell} & 0
    \end{pmatrix} \mathbf{a}_{\mathcal{L}}^{\text{out}},
\end{equation}
which describes the transfer of waves between connected vertices in the filter bonds and thus defines $P$. By substitution of (\ref{eq:transfer}) into equation (\ref{eq: Filter Scattering Matrix}) we define the filter scattering matrix as, 
\begin{equation}
    \mathbf{a}^{\text{out}}_{\mathcal{L}}= S_{\mu} \mathbf{a}^{\text{in}}_{\mathcal{L}}
 = \begin{pmatrix}
        r_{\mu} & t_{\mu} \\
        t_{\mu} & r_{\mu}
    \end{pmatrix} \mathbf{a}^{\text{in}}_{\mathcal{L}}
\end{equation}
where the lead transmission coefficient is
\begin{equation}\label{eq: Transmission Definition}
    t_{\mu} = \frac{i \sin(k \ell_{\mu})}{\cos ( \kappa_y \mu \ell - \Phi) - \cos(k \ell_{\mu})  + i \sin(k \ell_{\mu}) \left( 1 - \frac{\lambda}{2ik}\right)}
\end{equation}
with the reflection coefficient being trivially $r_{\mu} = t_{\mu} - 1$. With that we have derived the scattering properties of the filter and gained a new tuning parameter $\Phi$ is is a consequence of the introduction of the vector potential $A$ is the quantum graph. We will now show how this new parameter allows ideal transmission or arbitrary angles.

\section{Results: The Scattering Properties of the Filter}\label{sec: results}

The transmission profile of the filter has the rather odd property of being like a Kronecker-delta function scaled by some free parameter when at resonance. Formally,

\begin{widetext}
\begin{equation}\label{Binary Transmission}
    t_\mu = \frac{2ik}{2ik - \lambda} \delta_{k\ell_{\mu}, p\pi}\delta_{\kappa_y , \kappa_y^{(q)}}
    = \frac{2ik}{2ik - \lambda}
    \begin{cases}
    1, & \text{if } k\ell_{\mu} \rightarrow  p\pi \text{ and } \kappa_y = \kappa_y^{(q)} = \frac{\Phi + q\pi}{\mu\ell} \\
    0, & \text{if } k\ell_{\mu} \rightarrow p\pi \text{ and } \kappa_y \neq \kappa_y^{(q)}.
    \end{cases}
\end{equation}
\end{widetext}
Here, $p,q \in \mathbb{Z}$ and $\kappa_y^{(q)}$ represents a discrete set of tangential wave vectors. Note that the new amplitude scaling parameter $2ik/(2ik - \lambda)$ is equivalent to the transmission coefficient of a vertex of valency $v = 2$ as defined in (\ref{eq: Vertex Transmission and Reflection}); effectively completely removing the filter, decoupling each vertex and replacing it with a single vertex with variable transmission properties. Additionally we see when $\lambda = \Phi = 0$ we restore the solutions in the previous work on this topic \cite{lawrie2025nondiffracting} where one has unit transmission at discrete symmetrically-paired sets of incidence angles determined solely by the graph topology. By introducing non-zero $\lambda$ and $\Phi$ we see clearly that the discrete angle and amplitude of transmission can be tuned arbitrarily, resulting in a programmable steering device. See the filter in action in Figure \ref{fig: Results} for two beams incident on the filter with angles $\theta = 0$ and $\pi/4$ radians. To see how to couple discrete boundaries to semi-infinite scattering environments and construct beam solutions see \cite{lawrie2023engineering} and \cite{lawrie2022quantum} respectively.

\section{Conclusion}\label{sec: Conclusion}

In this work, we introduce a filtering device that enables discrete angular transmission at arbitrary angles with tunable amplitude. Using the scattering framework of quantum graph theory, we analytically investigate its properties. The device's graph forms a periodic interface of thin channels (edges) that couple boundary vertices in a non-trivial manner, supporting bound states that block energy transmission except at specific discrete angles. To tailor these angles for unit transmission, we incorporate a magnetic potential by configuring the graph bonds as a helix around a solenoid. The resulting potential breaks time-reversal symmetry, rendering the transmission angle a freely adjustable parameter. Furthermore, we control the transmission amplitude by applying a $\delta$-type boundary condition at the vertex connecting the filter to the environment; tuning the $\delta$ parameter makes the amplitude fully customizable. The outcome is a flux-tunable angular filter that supersedes topology-limited passbands with a programmable steering mechanism, expanding possibilities in wave filtering and beam shaping. We foresee applications in analogue wave computing and edge detection, offering advantages for medical imaging, non-destructive evaluation, remote sensing, and related fields.

\section*{Acknowledgements}
The authors acknowledge the financial support by the EPSRC in the form of the Postdoctoral Prize Fellowship (T.M.L., Grant No.
 EP/W524402/1) and Doctoral Training Partnership (O.M.B.). All data created during this research will be made available upon reasonable request to the corresponding author.

\bibliography{apssamp}% Produces the bibliography via BibTeX.

\end{document}